\documentclass[12pt]{iopart}
\usepackage{iopams}

\expandafter\let\csname equation*\endcsname\relax
\expandafter\let\csname endequation*\endcsname\relax
\usepackage{mathtools}
\usepackage{amssymb} 

\usepackage{dsfont}	
\usepackage{cases}
\usepackage{physics} 
\usepackage{easybmat}
\usepackage{cite}
\usepackage{hyperref}
\hypersetup{colorlinks}
\usepackage{textcomp}

\usepackage{xcolor}
\definecolor{cblue}{rgb}{0.16, 0.32, 0.75}
\definecolor{cred}{rgb}{0.7, 0.11, 0.11}
\hypersetup{%
	,linkcolor=cred
	,citecolor=cblue
	,urlcolor=gray
}

\usepackage{tikz}
\usetikzlibrary{decorations.markings, decorations.pathmorphing}
\pgfdeclarelayer{bg}   
\pgfsetlayers{bg,main}

\newcommand{\iu}{\mathrm{i}\mkern1mu}	
\DeclareMathOperator{\diag}{diag}
\newcommand{\id}{\mathrm{I}}

\begin{document}

	\title{Dimensional reduction of the Dirac theory}
	
	\author{
		Giuliano~Angelone$^{1,2}$,
		Elisa~Ercolessi$^{3,4}$,
		Paolo~Facchi$^{1,2}$, 
		Davide~Lonigro$^{1,2}$,
		Rocco~Maggi$^{1,2}$,
		Giuseppe~Marmo$^{5,6}$,
		Saverio~Pascazio$^{1,2}$, 
		Francesco~V.~Pepe$^{1,2}$
	}
	
	\address{$^1$ Dipartimento di Fisica, Universit\`a di Bari, I-70126 Bari, Italy}
	\address{$^2$ INFN, Sezione di Bari, I-70126 Bari, Italy}
	
	\address{$^3$ Dipartimento di Fisica e Astronomia, Universit\`a di Bologna, I-40127 Bologna, Italy}
	\address{$^4$ INFN, Sezione di Bologna, I-40127 Bologna, Italy}
	
	\address{$^5$ Dipartimento di Fisica, Universit\`a di Napoli Federico II, I-80126, Naples, Italy}
	\address{$^6$ INFN, Sezione di Napoli, I-80126, Naples, Italy}

	\begin{abstract}
		We perform a reduction from three to two spatial dimensions of the physics of a spin-\textonehalf{} fermion coupled to the electromagnetic field, by applying Hadamard's method of descent. We consider first the free case, in which motion is determined by the Dirac equation, and then the coupling with a dynamical electromagnetic field, governed by the Dirac-Maxwell equations. We find that invariance along one spatial direction splits the free Dirac equation in two decoupled theories. On the other hand, a dimensional reduction in the presence of an electromagnetic field provides a more complicated theory in $2+1$ dimensions, in which the method of decent is extended by using the covariant derivative. Equations simplify, but decoupling between different physical sectors occurs only if specific classes of solutions are considered. 	
	\end{abstract}
	

	\maketitle
	
	\section{Introduction}
	While the most natural setting of a physical theory in spacetime is a $(3+1)$-dimensional manifold, there is no obstruction, in principle, to formulating self-consistent theories in a higher- or lower-dimensional manifold~\cite{ehrenfest,hardy}. 	
	From a strictly physical point of view, these theories can provide a description of phenomena occurring in presence of an effective dimensional reduction, or, on the other hand, encode for convenience a theory in $3+1$ dimensions into a larger spacetime.
	
	Many partial differential equations of physics can be generalized to arbitrary spatial dimensions: this is notably the case for all problems where the Laplacian is the relevant differential operator acting on the spatial variables. In these cases, one can investigate how the solutions of equations appearing in the same form are affected by dimensionality, and possibly explore those properties that are peculiar to specific dimensions, $3+1$ in particular. A paramount example is the wave equation: around 1900, J.~Hadamard found out qualitative differences for propagation of a localized perturbation in odd and larger than 1 spatial dimensions, where perturbation concentrates around the wavefront, and even dimensions, characterized by the presence of a trailing edge following the wavefront~\cite{hadamard,ehrenulen,balasz,courant,evans}.
	
	Apart from fundamental reasons, the interest in low-dimensional theories is twofold. On one hand, they are of tantamount importance in the formulation of quantized field theories \cite{qed1,qed2,qed3,qed4,qed5,qed6}, on the other hand, the technological developments of the last few years have enabled us to engineer and control truly low-dimensional systems, yielding some fascinating dimension-dependent features~\cite{wheelerdirac,qsim_book,qsim1,qsim2,qsim3,qsim4,qsim5,qsim6,roy,wqed1,wqed2,graphene}. It is thus worth studying whether self-consistent low-dimensional theories can be obtained starting from a more familiar $(3+1)$-dimensional one. 
	
	There are two possible approaches to dimensional reduction. The first and most common one consists in formulating a lower dimensional version of the theory in $3+1$ dimensions, characterized by the same \textit{ab initio} properties of the starting point, as long as they are allowed by the new dimensionality \cite{lapidus,moreno,wheelerem,mcdonald,boito,GBB}. 
	An alternative approach is represented by the \textit{method of descent}. This tool was originally formulated by Hadamard to solve several evolution problems of classical physics in a generic number of spatial dimensions: according to his own words, ``it consists in noticing that he who can do more can do less''~\cite{hadamard}. The descent method represented the key to identify the aforementioned difference in the behavior of wave equation solutions for even and odd spatial dimensions. 
	
	While the descent method was conceived as a tool for solving equations, the underlying idea can be used in the reverse direction to perform the dimensional reduction of a given physical theory. Actually, a low-dimensional model can be considered as an instance of the theory where all the relevant quantities are uniform along one or more spatial directions. The low-dimensional equations of motion then follow directly from those in $3+1$ dimensions, by imposing such invariance. Reasonably enough, dimensional reduction by descent should yield a reduced version of the original $3+1$ theory among its products. However, the application of the descent method to electromagnetism~\cite{descent} shows us that this is not the whole story: besides the expected theory, other independent theories in the considered reduced dimensionality can emerge, sometimes with strikingly different physical properties.
		
	Following this line of research, in this work we shall apply the descent method to the equations of motion of a charged spin-\textonehalf\@ particle. We will start from the free Dirac theory, and then extend the results to the case of the minimal coupling with the electromagnetic field. After recalling, in Section~\ref{sec:prelim}, relevant notions about the Dirac equation in $3+1$ dimensions, we shall tackle in Section~\ref{sec:freedirac} the problem of dimensional reduction of the free Dirac equation, comparing the results with the natural generalization to an arbitrary dimension of the Dirac operator as a ``square root'' of the Klein-Gordon operator. 
	
	We will show that the method of descent provides two independent sectors, governed by non-interacting Dirac equation in $2+1$ dimensions. 
	Then, in Section~\ref{sec:diracem}, after briefly recalling the results obtained for the free electromagnetic field, we shall apply the same procedure to the coupled Dirac-Maxwell equations. We will observe that it is not trivial, in the case of interacting fields, to obtain a decoupled $2+1$ version of the Dirac theory by simply requiring invariance along one direction. The usual formulation of QED in $2+1$ dimensions emerges only by considering specific classes of solutions after dimensional reduction. An interesting spinoff of our analysis will be a nontrivial extension of Hadamard's descent method, with the introduction of the covariant derivative.
	
	\section{Preliminaries}\label{sec:prelim}

	The starting manifold of our analysis is the familiar $(3+1)$-dimensional Minkowski spacetime with metric tensor $(\eta^{\mu\nu})=\diag(+1,-1,-1,-1)$. Unless otherwise specified, Einstein summation convention is understood. In order to clearly distinguish between a $(3+1)$- and a $(2+1)$-dimensional context, three different kinds of indices will often be used: as common practice, Greek indices $\mu,\nu,\rho$, taking values in $\{0,1,2,3\}$, will refer to the $(3+1)$-dimensional coordinates, and  Latin indices $i, j, k$, with values in $\{1, 2, 3\}$, to its spatial coordinates. In addition,  Latin indices $a,b,c$, with values in $\{0,1,2\}$, will be reserved for the $(2+1)$-dimensional coordinates. This convention applies to free and contracted indices. Natural units will be adopted, and EM units will be rationalized.
	
	\subsection{Dirac equation in \texorpdfstring{$3+1$}{3+1} dimensions}\label{sec:normdirac}

	In a $(3+1)$-dimensional spacetime, the Dirac Lagrangian, describing a free spin-\textonehalf{} particle of mass $m$, reads
	\begin{equation}\label{eq:lagrDirac}
		\mathcal L=
		\overline \Psi\qty(\iu\gamma^\mu\partial_\mu-m \id_4)\Psi\, .
	\end{equation}
	Here, $\Psi=\Psi(x)$ is a four-component wavefunction, often called \emph{Dirac spinor} or \emph{bispinor}, $\id_4$ is the $4\times 4$ identity matrix (that will be kept implicit in most of the equations henceforth), $(\gamma^\mu)_{\mu=0,1,2,3}$ is a quadruple of $4\times 4$ matrices, usually referred to as the \emph{gamma matrices}, satisfying the anticommutation property
	\begin{equation}\label{eq:cliff}
		\acomm{\gamma^\mu}{\gamma^\nu}=2\eta^{\mu\nu}\id_4\,,
	\end{equation}
	and $\overline\Psi=\Psi^\dagger\gamma^0$ is the \emph{Dirac adjoint spinor}. 
	By requiring that
	\begin{equation}\label{eq:hermigamma}
		{\gamma^\mu}^{\dagger}=\gamma^0\gamma^\mu\gamma^0\,,
	\end{equation}
	one gets that $\gamma^0$ is Hermitian (hence $\overline\Psi=(\gamma^0\Psi)^\dagger$), and the $\gamma^i$'s  are anti-Hermitian, and the action associated to the Lagrangian $\mathcal L$ is real-valued.
	It is also useful to introduce a fifth matrix
	\begin{equation}
		\gamma^5=\iu\gamma^0\gamma^1\gamma^2\gamma^3 ,
	\end{equation}
	which is Hermitian, involutive, and anticommuting with all the $\gamma^\mu$'s. The matrices $\id_4$, $(\gamma^\mu)$, $(\gamma^\mu \gamma^\nu)_{\mu<\nu}$, $(\gamma^5 \gamma^\mu )$, and $\gamma^5$ form a basis of the space of $4\times 4$ complex matrices~\cite{Go55}. The anticommutation relations~\eqref{eq:cliff} have the following immediate consequences:  the gamma matrices are traceless, $\gamma^0$ being involutive, and each $\gamma^i$ anti-involutive (therefore, the condition~\eqref{eq:hermigamma} is equivalent to assuming the gamma matrices to be unitary).
	
	The stationary points of the action corresponding to the Dirac Lagrangian~\eqref{eq:lagrDirac} are determined by the Dirac equation~\cite{dirac,thaller}
	\begin{equation}\label{eq:Dirac}
		\qty(\iu\gamma^\mu\partial_\mu-m\id_4)\Psi=0\,.
	\end{equation}
	The four-component differential operator $(\iu\gamma^\mu\partial_\mu-m\id_4)$ applied to the Dirac spinor squares to the Klein-Gordon operator, in the following sense:
	\begin{equation}
		(\iu\gamma^\mu\partial_\mu-m\id_4)^\dagger\,(\iu\gamma^\mu\partial_\mu-m\id_4)=\left(\partial_\mu\partial^\mu+m^2 \right)\id_4 .
	\end{equation}
	The matrices $S^{\mu\nu}=\tfrac{\iu}{4}[\gamma^\mu,\gamma^\nu]$ generate the $(\tfrac{1}{2},0)\oplus(0,\tfrac{1}{2})$ representation of the restricted Lorentz group. 
	In particular, by introducing the six Hermitian matrices  
	\begin{equation}\label{eq:alphabeta}
		\Sigma^i=\tfrac{\iu}{2}\epsilon^{ijk}\gamma^j\gamma^k, \qquad \alpha^i=\gamma^0\gamma^i \,,
	\end{equation}		
	a proper rotation is represented by $\exp(-\tfrac{\iu}{2}\vb*\theta\cdot\vb*\Sigma)$, and a boost by $\exp(-\tfrac{1}{2}\vb*\eta\cdot\vb*\alpha)$. The Noether current  
	\begin{equation}\label{eq:jmuvett}
		(j^\mu)=\qty(\overline{\Psi}\gamma^\mu\Psi)=\qty(\Psi^\dagger\Psi,
		\Psi^\dagger\vb*\alpha\Psi)\,,
	\end{equation}
	related to the $\mathrm{U}(1)$ symmetry $\Psi\to\mathrm{e}^{\iu\theta}\Psi$ of the Lagrangian~\eqref{eq:lagrDirac}, is manifestly covariant, behaving as a vector under restricted Lorentz transformations, and can be interpreted as a probability density four-current.
	
	Distinct families of matrices satisfying Eq.~\eqref{eq:cliff} correspond to different \textit{representations} of the Dirac spinors and the Dirac algebra, all of them being linked via unitary transformations $U\in\mathrm{U}(4)$,
	\begin{equation}
		\Psi\to \Psi'=U\Psi, \qquad \gamma^{\mu}\to {\gamma'}^{\mu}=U\gamma^{\mu}U^{\dagger} ,
	\end{equation}
	preserving both the anticommutation relations~\eqref{eq:cliff} (as a more general similarity transformation would do) and the conditions~\eqref{eq:hermigamma}~\cite{Go55}. In the following, different representations will be labelled by specific labels on the spinor and the gamma matrices. Dirac's original choice,
	\begin{equation}\label{eq:alphabetaDirac}
		\gamma_{\mathrm{D}}^0
		=\mqty(\id_2& 0\\ 0&-\id_2)	, \qquad
		\gamma_{\mathrm{D}}^i
		=\mqty( 0&\sigma^{i}\\ -\sigma^{i}&0 ) , \qquad
		\Psi_{\mathrm{D}}=\mqty(\phi\\		\chi) ,
	\end{equation}	
	where
	\begin{equation}
		\sigma^1 = \mqty( 0 & 1 \\ 1 & 0), \qquad \sigma^2 = \mqty( 0 & -\iu \\ \iu & 0), \qquad \sigma^3 = \mqty( 1 & 0 \\ 0 & -1)
	\end{equation}
	are the Pauli matrices,
	is known as the \emph{standard} or \emph{Dirac} \emph{representation}. In this representation, the generators~\eqref{eq:alphabeta}
	take the simple forms 	\begin{equation}\label{eq:alphasigmaSR}
		\vb*\Sigma_{\mathrm {D}}
		=\mqty( 		\vb*\sigma&0\\ 0 &		\vb*\sigma), \qquad
		\vb*\alpha_{\mathrm {D}}
		=\mqty(  0 &		\vb*\sigma\\ 		\vb*\sigma&0).
	\end{equation}	
	In particular, it is worth noticing that the spin operator along the $z$ direction, $\Sigma_{\mathrm{D}}^3$, is diagonal in the Dirac representation. 
	
	\subsection{Dirac equation in arbitrary spatial dimensions}
	
	The anticommutation relations~\eqref{eq:cliff} can be imposed in a Min\-kowski spacetime of arbitrary spatial dimensionality $n$, starting from  $n+1$ square matrices $(\Gamma^A)_{0\le A\le n}$, of order $N=2^{\lfloor(n+1)/2\rfloor}$ (with $\lfloor x\rfloor$ denoting the integer part of $x$), satisfying the anticommutation algebra $\acomm{\Gamma^A}{\Gamma^B}=2\eta^{AB}\id_{N}$~\cite{Go55,BrauerWeyl}, with 
	$(\eta^{AB})=\diag(+1,-1,\dots,-1)$, and $\id_N$ being the $N\times N$ identity matrix. 
	Therefore, the free Dirac equation can be generalized to an arbitrary number of spatial dimensions, a feature essentially noticed from the beginning by Dirac himself \cite{history}. A ``natural'' notion of an $(n+1)$-dimensional Dirac theory can be achieved by following the same pattern as in Subsection~\ref{sec:normdirac}. Such a theory will be formulated in terms of $N$-dimensional objects, with $(\iu \Gamma^A\partial_A-m I_N)^{\dagger}(\iu \Gamma^A\partial_A-m I_N)$ equaling the Klein-Gordon operator in $n+1$ dimensions.
	
	Here, we shall focus on the case of $2$ spatial dimensions, so the triple of $2\times 2$ gamma matrices $(\Gamma^a)$ will satisfy 
	\begin{align}\label{eq:cliffLD}
		\acomm{\Gamma^a}{\Gamma^b}=2\eta^{ab}\id_2, \qquad {\Gamma^a}^{\dagger} = \Gamma^0 \Gamma^a \Gamma^0 .
	\end{align}
	The corresponding Dirac Lagrangian $\overline{\psi}\qty(\iu\Gamma^a\partial_a-m\id_2)\psi$, with $\psi$ a two-component function of three variables and $\overline{\psi}=\psi^{\dagger}\Gamma^0$, yields the Euler-Lagrange equation
	\begin{equation}\label{eq:DiracLow}
		\qty(\iu\Gamma^a\partial_a-m\id_2)\psi=0\,,
	\end{equation}
	which will be referred to as the $(2+1)$-dimensional Dirac equation. One of the possible choices of $(2+1)$-dimensional gamma matrices in $2+1$ dimensions is represented by $(\sigma^3, \iu\sigma^2,-\iu\sigma^1)$~\cite{thaller}. 
	
	\section{Dimensional reduction of the free Dirac theory}\label{sec:freedirac} 
	We will perform the descent on the Dirac equation from $3$ to $2$ spatial dimensions along the $z$ coordinate, but any other direction would lead to analogous results. We expect a sensible $(2+1)$-dimensional free Dirac theory 
	to be covariant with respect to the $(2+1)$-dimensional Lorentz group $\mathrm{O}(1,2)$ of the $z$-preserving transformations
	\begin{align}\label{eq:lambda}
		\varLambda	
		=\left(\begin{BMAT}{c1c}{c1c}
			\begin{BMAT}[5pt]{c}{c} L \end{BMAT} & \\
			&\,Q\,
		\end{BMAT}\right)\,,
	\end{align}
	with $L \in\mathrm{O}(1,2)$, and $Q \in\mathrm{O}(1)=\{+1,-1\}$. In a $(2+1)$-dimensional context, attributes such as ``covariant'', ``vector'', ``scalar'' will always be referred to this group or one of its subgroups. 
	
	Therefore, we specialize the Dirac differential problem into a $z$-in\-de\-pen\-dent one by imposing the \emph{descent condition}
	\begin{equation}
		\partial_3\Psi=0\,,
	\end{equation}
	which is both re\-pre\-sen\-ta\-tion-in\-de\-pen\-dent, and covariant.
	We end up with the ``reduced Dirac equation''
	\begin{equation}\label{eq:descent}
		\qty(\iu \gamma^a \partial_a-m\id_4)	\Psi=0\,,
	\end{equation}
	that can be regarded as the Euler-Lagrange equation of the ``reduced Lagrangian''
	\begin{equation}\label{eq:lagrDesc}
		\tilde{\mathcal L}=
		\overline \Psi\qty(\iu\gamma^a\partial_a-m\id_4)\Psi\,.
	\end{equation}
	
	As we will show in the following, Eq.~\eqref{eq:descent} is equivalent to the pair of  uncoupled equations:
	\begin{align}
		\qty(\iu\gamma_{+}^{a}\partial_a-m\id_4)\Psi_{+}=0\,,\label{eq:Gammap}\\
		\qty(\iu\gamma_{-}^{a}\partial_a-m\id_4)\Psi_{-}=0\,,\label{eq:Gammam}
	\end{align}
	where $(\gamma_{+}^{a})_{0\le a\le 2}$ and $(\gamma_{-}^{a})_{0\le a\le 2}$ are families of $4\times 4$ matrices satisfying 
	\begin{equation}\label{cliffa}
		\acomm{\gamma_{\pm}^a}{\gamma_{\pm}^b}=2\eta^{ab}P_\pm\,,
		\qquad
		\gamma^a_\pm P_\pm=\gamma^a_\pm=P_\pm\gamma^a_\pm\,,
	\end{equation}
	with $P_+$ and $P_-$ two complementary orthogonal projections of rank 2, thus satisfying
	\begin{equation}\label{eq:projIdem}
		P_+ P_-=0, \qquad P_{\pm}^2 = P_{\pm}^{\dagger} = P_{\pm}  , \qquad P_++P_-=\id_4 \,.
	\end{equation}
	
	Moreover, Eqs.~\eqref{eq:Gammap}--\eqref{eq:Gammam} turn out to be $(2+1)$-dimensional Dirac equations, that can be explicitly put  in the form of Eq.~\eqref{eq:DiracLow} in a suitable basis, such that $P_+=\id_2 \oplus\, 0_2$ and $P_-= 0_2 \oplus \id_2$.	
	Notice that, despite the formal analogy between the equations, dimensional reduction generates \textit{two independent Dirac theories} in (2+1) dimensions. The derivation of two independent dimensionally-reduced models by descent was already found in the electromagnetic case, as we shall outline in the following. There, however, the theories are also formally different.
	
	In Section~\ref{sec:sr}, we shall first discuss the problem in the standard representation (which, quite by chance, turns out to be particularly advantageous to perform the descent along $z$), and then formulate in Subsection~\ref{sec:alg} a general proof of the above statement that is independent of the representation. 
	A discussion of these results is finally given in Section~\ref{sec:remarks}.
	
	\subsection{Descent in the standard representation}\label{sec:sr}
	By conveniently labelling the spinor components in the standard representation as
	\begin{align}\label{eq:spinoSR}
		\Psi_{\mathrm{D}}=\mqty(\phi\\		\chi)
		=\mqty( \phi_\uparrow\\	\phi_\downarrow\\ \chi_\downarrow\\ \chi_\uparrow)
		\,,
	\end{align}
	the Dirac equation~\eqref{eq:Dirac} has the matrix form
	\begin{align}
		\begin{pmatrix}
			\iu\partial_0-m&0&\iu\partial_3&\iu\partial_1+\partial_2\\
			0&\iu\partial_0-m&\iu\partial_1-\partial_2&-\iu\partial_3\\
			-\iu\partial_3&-\iu\partial_1-\partial_2&-\iu\partial_0-m&0\\
			-\iu\partial_1+\partial_2&\iu\partial_3&0&-\iu\partial_0-m
		\end{pmatrix}
		\begin{pmatrix}
			\phi_\uparrow\\	\phi_\downarrow\\	\chi_\downarrow\\	\chi_\uparrow
		\end{pmatrix}=0\,,\label{eq:diracDirac}
	\end{align}
	equivalent to the system of four equations
	\begin{equation}
		\begin{cases}
			(\iu\partial_0-m)	\phi_\uparrow+\iu\partial_3\chi_\downarrow+ (\iu\partial_1+\partial_2)\chi_\uparrow=0\,, 
			\\
			(\iu\partial_0-m)\phi_\downarrow+(\iu\partial_1-\partial_2)\chi_\downarrow-\iu\partial_3\chi_\uparrow=0\,,	
			\\
			(\iu\partial_0+m)\chi_\downarrow	+\iu\partial_3\phi_\uparrow+(\iu\partial_1+\partial_2)\phi_\downarrow
			=0\,,	\label{eq:FreeIII}\\
			(\iu\partial_0+m)\chi_\uparrow+
			(\iu\partial_1-\partial_2)\phi_\uparrow-\iu\partial_3\phi_\downarrow=0\,. 
		\end{cases}
	\end{equation}
	Notice that all the coupling terms between $\uparrow$ and $\downarrow$ components are in the form of a derivative $\partial_3$ with respect to $z$.

	Consequently, the condition $\partial_3\Psi_\mathrm{D}=0$ breaks all and only the couplings between the $\uparrow$ and $\downarrow$ components, so that the Dirac equation takes the matrix form
	\begin{equation}\label{eq:diracDiracz}
		\left(\begin{array}{@{}c|cc|c@{}}
			\iu\partial_0-m&0&0&\iu\partial_1+\partial_2\\
			\hline
			0& \iu\partial_0-m& \iu\partial_1-\partial_2&0\\
			0&-\iu\partial_1-\partial_2&-\iu\partial_0-m&0\\
			\hline
			-\iu\partial_1+\partial_2&0&0&\-\iu\partial_0-m
		\end{array}\right)
		\left(\begin{array}{@{}c@{}}
			\phi_\uparrow\\
			\hline
			\phi_\downarrow\\
			\chi_\downarrow\\
			\hline
			\chi_\uparrow
		\end{array}\right)=0\,,
	\end{equation}
	and the wavefunction is partitioned into two independently evolving sectors, respectively spanned by the $\uparrow$ components, whose dynamics is determined by
	\begin{equation}
		\begin{cases}
			(+\iu\partial_0-m)	\phi_\uparrow+(+\iu\partial_1+\partial_2)\chi_\uparrow=0\,, 
			\\
			(-\iu\partial_0-m)\chi_\uparrow+(-\iu\partial_1+\partial_2)\phi_\uparrow=0\,,\label{eq:descFreeII}
		\end{cases}
	\end{equation}
	and the $\downarrow$ components, obeying the equations
	\begin{equation}
		\begin{cases}
			(+\iu\partial_0-m)\phi_\downarrow+(+\iu\partial_1-\partial_2)\chi_\downarrow=0\,,
			\\
			(-\iu\partial_0-m)\chi_\downarrow+(-\iu\partial_1-\partial_2)\phi_\downarrow=0\,.\label{eq:descFreeIV}
		\end{cases}
	\end{equation}
	The only difference between the $\uparrow$ and $\downarrow$ equations consists in the sign before the $\partial_2$ terms, which entails that the subsystems~\eqref{eq:descFreeII} and~\eqref{eq:descFreeIV} are transformed into each other by a reflection of the $y$-axis $(t,x,y,z)\to(t,x,-y,z)$.
	
	At this point, by observing that both the triples of $2 \times 2$ matrices
	\begin{align}\label{GammaD}
		\qty(\Gamma_{\uparrow}^a)=(\sigma^3,\iu\sigma^2,-\iu\sigma^1)	\,,
		&&\qty(\Gamma_{\downarrow}^a)=(\sigma^3,\iu\sigma^2,+\iu\sigma^1)\,,
	\end{align}
	satisfy the anticommutation relations~\eqref{eq:cliffLD},
	and by introducing the two-components wavefunctions
	\begin{equation}\label{eq:Dspin}
		\psi_{\uparrow}
		=\begin{pmatrix}
			\phi_\uparrow\\		\chi_\uparrow
		\end{pmatrix}\,,
		\qquad
		\psi_{\downarrow}
		=\begin{pmatrix}
			\phi_\downarrow\\\chi_\downarrow
		\end{pmatrix}\,,
	\end{equation}
	the subsystems~\eqref{eq:descFreeII} and~\eqref{eq:descFreeIV} can be rewritten as 
	\begin{align}
		\qty(\iu\Gamma_{\uparrow}^{a}\partial_a-m\id_2)\psi_{\uparrow}=0\,,\label{eq:Gammaup}\\
		\qty(\iu\Gamma_{\downarrow}^{a}\partial_a-m\id_2)\psi_{\downarrow}=0\,,\label{eq:Gammadown}
	\end{align}
	namely, as a pair of $(2+1)$-dimensional Dirac equations. 
	
	The partition of the spinor in Eq.~\eqref{eq:diracDiracz} can be easily interpreted by looking at the generators of $\mathrm{SO^+(1,2)}$, which are proportional to the matrices
	\begin{align}
		\Sigma^3_{\mathrm D}=
		\left(\begin{BMAT}[1pt]{c1cc1c}{c1cc1c}
			1&&\phantom{-1}&\\
			\phantom{-1}&-1&&\\
			&&1&\\
			&&&-1
		\end{BMAT}\right),
		&&
		\alpha^1_{\mathrm D}=
		\left(\begin{BMAT}[1pt]{c1cc1c}{c1cc1c}
			&&\phantom{-1}&\iu\\
			&\phantom{-1}&\iu&\\
			\phantom{-1}&\iu&&\\
			\iu&&&\phantom{-1}
		\end{BMAT}\right),
		&&
		\alpha^2_{\mathrm D}
		=\left(\begin{BMAT}[1pt]{c1cc1c}{c1cc1c}
			&&&1\\
			&\phantom{-1}&-1&\\
			&1&&\phantom{-1}\\
			-1&&&
		\end{BMAT}\right).
	\end{align}
	Actually, the transformation generated by $(\Sigma_{\mathrm{D}}^3,\alpha_{\mathrm{D}}^1,\alpha_{\mathrm{D}}^2)$ does not mix the $\uparrow$ and $\downarrow$ components: our partition thus coincides with the decomposition of the spinor with respect to this group, satisfying the expected $(2+1)$-dimensional covariance.
	
	It is worth remarking that the pair~\eqref{eq:Gammaup}--\eqref{eq:Gammadown} of uncoupled Dirac equations emerges naturally in the  low-energy  description  of graphene  (see  e.g.\  Eq.~(19)  of~\cite{graphene}),  which  represents  a prototypical example of ($2+1$)-dimensional \textit{Dirac matter}.
	
	\subsection{Descent in a representation-independent approach}\label{sec:alg}
	The calculations in the Dirac representation are facilitated by the fact that the generator $\Sigma_3$, representing the spin component along the descent direction $z$, is diagonal. Had we chosen a different representation, or had we performed the descent in the same representation but along a different direction, the resulting decoupling would have been less evident: of course we would have ended with two independently evolving 2-dimensional subspaces, but in general their vectors would have been written in terms of linear combinations of the spinor components. Therefore, it is worth reviewing the decoupling mechanism from a re\-pre\-sen\-ta\-tion-in\-de\-pen\-dent point of view.
	
	We can say that a proper linear
	decoupling consists in the existence of a pair of projections on nontrivial complementary subspaces, namely two operators $P_\pm$ satisfying the properties~\eqref{eq:projIdem}, with the possible exception of Hermiticity,
	acting exclusively on the spinorial degrees of freedom of $\Psi$, such that the projected wavefunctions $P_+\Psi$ and $P_-\Psi$ are mutually decoupled in the \textit{reduced} Dirac equation~\eqref{eq:descent}. The latter can be decomposed, by means of the identity resolution provided by $P_{\pm}$, into the pair
	\begin{align}
		P_+(\iu\gamma^a\partial_a-m)P_+\Psi+\iu P_+\gamma^a P_-\partial_a\Psi&=0\,,\\
		P_-(\iu\gamma^a\partial_a-m)P_-\Psi+\iu P_-\gamma^a P_+\partial_a\Psi&=0\,.
	\end{align}
	
	A necessary and sufficient condition for $P_+\Psi$ and $P_-\Psi$ to be decoupled is the vanishing of the coupling terms $\iu P_+\gamma^a P_-\partial_a\Psi$ and $\iu P_-\gamma^a P_+\partial_a\Psi$. Considering the property $P_+P_-=P_-P_+=0$ and the arbitrariness of the spinor derivatives, these conditions are equivalent to 
	\begin{equation}\label{eq:projCommuto}
		\left[P_+,\gamma^a\right]=0 \qquad \text{for } a=0,1,2.
	\end{equation}
	
	In order to determine the $4\times 4$ complex matrices $P_{\pm}$, we decompose them in the form
	\begin{equation}\label{eq:decomposition}
		P_{\pm}=a_0\id+\sum_\mu b_\mu\gamma^\mu+ \sum_{\mu<\nu}c_{\mu\nu}\gamma^\mu\gamma^\nu+\sum_\mu d_\mu\gamma^\mu\gamma^5 + a_5 \gamma^5\,,
	\end{equation}
	where we have temporarily dropped Einstein's convention, yielding the commutators
	\begin{equation}
		\tfrac{1}{2}[P_{\pm},\gamma^a]=\sum_{\mu\ne a} b_\mu\gamma^\mu\gamma^a+\eta^{aa}\biggl(- \sum_{a<\nu}c_{a\nu}\gamma^\nu
		+\sum_{\mu<a}c_{\mu a}\gamma^\mu- d_a\gamma^5\biggr)- a_5 \gamma^a\gamma^5\,.
	\end{equation}
	By imposing the commutation conditions~\eqref{eq:projCommuto}, while letting the $(2+1)$-dimensional index $a$ take the values $0,1,2$, all the coefficients but $d_3$ must vanish, so $P_{\pm}=a_0\id+d_3\gamma^3\gamma^5$. Finally, by imposing the idempotence condition~\eqref{eq:projIdem},
	\begin{equation}
		a_0\id+d_3\gamma^3\gamma^5=\qty(a_0^2+{d_3}^2)\id + 2 a_0 d_3 \gamma^3\gamma^5\,,
	\end{equation}
	we get $a_0=1/2$ and $d_3=\pm1/2 $, and the projections are finally obtained:
	\begin{equation}\label{eq:Ppm}
		P_\pm=\frac{\id \pm \gamma^3\gamma^5}{2}\,.
	\end{equation}
	We are thus led to consider the matrix
	\begin{equation}
		\kappa^3= P_+ - P_- =\gamma^3\gamma^5=\iu\gamma^0\gamma^1\gamma^2
		=\gamma^0\Sigma^3\,,
	\end{equation}
	which is traceless, because $\mathrm{tr}\qty(\gamma^3\gamma^5)=-\mathrm{tr}\qty(\gamma^5\gamma^3)$, as well as Hermitian and involutive, since it is the product of the commuting Hermitian and involutive matrices $\gamma^0$ and $\Sigma^3$. Therefore, its eigenvalues are $1$ and $-1$, each with multiplicity $2$, corresponding to the \emph{orthogonal} projections $P_+$ and $P_-$, respectively. Moreover, by construction we have $[\gamma^a,\kappa^3]=0$.

	Now, by setting
	\begin{equation}\label{eq:Decomp}
		\Psi_\pm=P_\pm\Psi, \qquad
		\gamma^a_\pm=P_\pm\gamma^a P_\pm, 
	\end{equation}
	the following decompositions hold:
	\begin{align}
		\Psi&=\Psi_+ +\Psi_-\,,\label{eq:PsiDecomp}\\
		\gamma^a&=\gamma^a_+ +\gamma^a _-\,,\label{eq:gammaADecomp}
	\end{align}
	and the reduced equation~\eqref{eq:descent} can be written as the pair of uncoupled equations
	\begin{align}
		\qty(\iu\gamma_{+}^{a}\partial_a-m\id_4)\Psi_{+}=0\,,\label{eq:gammap}\\
		\qty(\iu\gamma_{-}^{a}\partial_a-m\id_4)\Psi_{-}=0\,,\label{eq:gammam}
	\end{align}
	where $(\gamma_{\pm}^a)$ are two sets  matrices satisfying 
	\begin{equation}\label{eq:clifflike}
		\acomm{\gamma^a_\pm}{\gamma^b_\pm}=\acomm{P_\pm\gamma^a P_\pm}{P_\pm\gamma^b P_\pm}=2\eta^{ab}P_\pm\,.
	\end{equation}
	
	At this point, we can show that that Eqs.~\eqref{eq:gammap}--\eqref{eq:gammam} are $(2+1)$-dimensional Dirac equations, in the following sense. 
	First, we can always choose a representation where the projections are diagonalized in the form 
	\begin{equation}\label{eq:projrepr}
		\hat P_{+}=\mqty(\id_2& 0\\0& 0 )\,,	
		\qquad
		\hat P_{-}=\mqty(0& 0\\0&\id_2)\,.
	\end{equation}
	According to the decompositions~\eqref{eq:PsiDecomp}--\eqref{eq:gammaADecomp}, $\hat\gamma^a$ and $\hat\Psi$, the expressions of the $a$-th gamma matrix and of the wavefunction in the chosen representation, have the block structure 
	\begin{equation}\label{eq:diagonal_decomp}
		\hat\gamma^a=\mqty(\hat\Gamma^a_+ & 0\\0&\hat\Gamma^a_-)\,, \qquad \hat\Psi=\mqty(\hat\psi_+\\ \hat\psi_-)\,.
	\end{equation}
	In this regard, notice that the existence of a projection commuting with each $\gamma^a$, in turn a necessary and sufficient condition for the decoupling of the reduced Dirac equation, is also equivalent to the simultaneous block-diagonalizability of the $(\gamma^a)$. 
	
	Then the sets $\bigl(\hat\Gamma^a_\pm\bigr)$ inherit their anticommutation relations from those of $\bigl(\hat\gamma^a_\pm\bigr)$ in Eq.~\eqref{eq:clifflike}, and turn out to be $(2+1)$-dimensional gamma matrices. The decoupled equations~\eqref{eq:gammap}--\eqref{eq:gammam} are equivalent to
	\begin{align}
		\bigl(\iu\hat\Gamma_{+}^{a}\partial_a-m\id_2\bigr)\hat\psi_{+}=0\,,\label{eq:hatGammap}\\
		\bigl(\iu\hat\Gamma_{-}^{a}\partial_a-m\id_2\bigr)\hat\psi_{-}=0\,,\label{eq:hatGammam}
	\end{align}
	which are in the form~\eqref{eq:DiracLow}, hence $(2+1)$-dimensional Dirac equations, defined on the two-dimensional images of $P_+$ and $P_-$, respectively.
	
	\subsection{Discussion of the results}\label{sec:remarks}

	\subsubsection{Reconnecting the pictures.}
	
	In order to check the consistency between the results of the last two subsections, we write $\kappa^3$ in the standard representation. Since 
	\begin{equation}
		\kappa^3_\mathrm{D} = \gamma_{\mathrm{D}}^3 \gamma_{\mathrm{D}}^5 =\diag(1,-1,-1,1) ,    
	\end{equation}
	its eigenspaces corresponding to $1$ or $-1$ are exactly the subspaces spanned by the $\uparrow$ and $\downarrow$ spinor components, respectively. Accordingly, Eqs.~\eqref{eq:Gammaup}--\eqref{eq:Gammadown} and Eqs.~\eqref{eq:gammap}--\eqref{eq:gammam} are just different expressions (in terms of algebraic objects of different orders, $2$ and $4$, respectively) of the same couple of $(2+1)$-dimensional Dirac equations.
	
	\subsubsection{Properties of the decoupling.} 
	
	First of all, we check that the decomposition underlying our decoupling, as well as our reduced equations are covariant with respect to the $(2+1)$-dimensional  Lorentz group~\eqref{eq:lambda}. Indeed, let us recall that $\gamma^5$ anticommutes with all the gamma matrices, hence commuting with all the generators of the restricted Lorentz group. As a consequence, we observe that $\kappa^3$ is left invariant by the restricted Lorentz transformations of the form~\eqref{eq:lambda}:
	\begin{align}
		D(\varLambda^{-1})\kappa^3 D(\varLambda)=	
		D(\varLambda^{-1})\gamma^3	D(\varLambda)
		D(\varLambda^{-1})\gamma^5	D(\varLambda)
		={\varLambda^3}_\mu\gamma^\mu\gamma^5
		=Q\gamma^3\gamma^5
		=\kappa^3\,,
	\end{align}
	for $Q=+1$, with $D$ denoting the $(\tfrac{1}{2},0)\oplus(0,\tfrac{1}{2})$ representation of the restricted Lorentz group.
	
	The reduction of the original equation in a pair of decoupled $(2+1)$-dimensional Dirac equations can also be interpreted at the level of Lagrangians as follows: the reduced Lagrangian~\eqref{eq:lagrDesc} can be written, without any further assumption, as the sum of two independent Lagrangians,
	\begin{equation}\label{eq:lagrDescDecomp}
		\tilde{\mathcal L}
		=\overline \Psi_+\qty(\iu\gamma_+^a\partial_a-m)\Psi_+ 
		+ \overline \Psi_-\qty(\iu\gamma_-^a\partial_a-m)\Psi_-
		=\tilde{\mathcal L}_+ +\tilde{\mathcal L}_-\,,
	\end{equation}
	with Eqs.~\eqref{eq:gammap}--\eqref{eq:gammam} being their respective Euler-Lagrange equations.

	\subsubsection{``Block-diagonal'' representations.}	
	
	Let us try to decompose $\gamma^3$ in a similar way as Eq.~\eqref{eq:gammaADecomp}.
	Its behaviour with respect to the projections $P_\pm$,
	\begin{align}
		\gamma^3 P_\pm=P_\mp\gamma^3\,,
	\end{align}
	yields the decomposition
	\begin{align}\label{eq:gamma3Decomp}
		\gamma^3=\gamma^3_{+,-}+\gamma^3_{-,+}\,,
	\end{align}
	where we have set
	\begin{align}\label{eq:gamma3PM}
		\gamma^3_{\pm,\mp}=P_\pm\gamma^3 P_\mp=\gamma^3 P_\mp=P_\pm\gamma^3\,.
	\end{align}
	The decompositions~\eqref{eq:gammaADecomp} and~\eqref{eq:gamma3Decomp} of the gamma matrices extend to the vector current~\eqref{eq:jmuvett}, since by setting
	\begin{align}\label{eq:japm}
		j_\pm^a&=\Psi^\dagger_\pm\gamma^0_\pm\gamma^a_\pm\Psi_\pm\,,\\
		j_{\pm,\mp}^3&=\Psi^\dagger_\pm\gamma^0_\pm\gamma^3_\mp\Psi_\mp\,.
	\end{align}
	we can accordingly write
	\begin{align}
		\overline{\Psi}\gamma^a\Psi&=j_+^a+j_-^a\,,\label{eq:currdecompa}\\
		\overline{\Psi}\gamma^3\Psi&=j_{+,-}^3+j_{-,+}^3\,.\label{eq:currdecomp3}
	\end{align}
	In matrix terms, Eq.~\eqref{eq:gamma3Decomp} tells us that, in the representation~\eqref{eq:projrepr}, where each $\gamma^a$ is block-diagonal, $\gamma^3$ has to be off-block-diagonal, specifically:
	\begin{align}
		\hat\gamma^3=\mqty(0 & \hat\Gamma^3_{+,-}\\\hat\Gamma^3_{-,+}&0 )\,.
	\end{align}
	
	The decomposition, determined by the descent, of the free Dirac theory into two independent models is thus clearly visualized in any representation such that the first three Dirac matrices are block diagonal. The Dirac equation takes the form:
	\begin{align}
		\left(\begin{BMAT}[4pt]{cc}{cc}
			\iu\hat\Gamma_{+}^{a}\partial_a-m\id_2&\iu\hat\Gamma_{+,-}^{3}\partial_3\\
			\iu\hat\Gamma_{-,+}^{3}\partial_3&\iu\hat\Gamma_{-}^{a}\partial_a-m\id_2
		\end{BMAT}\right)
		\left(\begin{BMAT}[4pt]{c}{cc}\hat\psi_+\\\hat\psi_-\end{BMAT}\right)=0\,.
	\end{align}		
	In this way, since terms proportional to $\partial_3$ are cancelled by the descent condition, the reduced equation~\eqref{eq:descent} can be recast in the block-diagonal form
	\begin{align}
		\left(\begin{BMAT}[4pt]{c1c}{c1c}
			\iu\hat\Gamma_{+}^{a}\partial_a-m\id_2 & 0\\
			0 & \iu\hat\Gamma_{-}^{a}\partial_a-m\id_2
		\end{BMAT}\right)
		\left(\begin{BMAT}[4pt]{c}{c1c}
			\hat\psi_+\\ 
			\hat\psi_-\end{BMAT}\right)=0\,.
	\end{align}		
	and the decoupling into two $(2+1)$-dimensional Dirac equation manifestly emerges.
	
	The computations in Section~\ref{sec:sr} can be used to single out a particular ``block-diagonal representation'', that can be defined through the unitary transformation relating it to the standard representation. By permuting the standard components as
	\begin{equation}\label{eq:perm}
		\hat\Psi=U\Psi_{ \mathrm{D}}
		=\left(\begin{BMAT}[1pt]{c}{cccc}
			\phi_\uparrow\\ \chi_\uparrow\\ 
			\phi_\downarrow\\ \chi_\downarrow	\end{BMAT}\right)\,, \qquad
		U=		\left(\begin{BMAT}[1pt]{cccc}{cccc}
			\vphantom{\phi_\uparrow}\mathmakebox[\widthof{$\phi_\uparrow$}]{1}&0&0&0\\ 	0&0&0&\vphantom{\chi_\uparrow}\mathmakebox[\widthof{$\chi_\uparrow$}]{1}\\ 	0&\vphantom{\phi_\downarrow}\mathmakebox[\widthof{$\phi_\downarrow$}]{1}&0&0\\ 	0&0&\vphantom{\chi_\downarrow}\mathmakebox[\widthof{$\chi_\downarrow$}]{1}&0
		\end{BMAT}\right)\,,
	\end{equation}
	the corresponding gamma matrices, $\hat\gamma^\mu =U\gamma_{\mathrm{D}}^\mu U^{\dagger}$, explicitly read
	\begin{align}\label{eq:Diracfreeunci}
		\hat\gamma^0=\mqty( \sigma^3&0 \\ 0&\sigma^3)\,,
		&& 
		\hat\gamma^1=\mqty( \iu\sigma^2&0 \\ 0&\iu\sigma^2)\,,
		&&
		\hat\gamma^2=\mqty( -\iu\sigma^1&0 \\ 0&\iu\sigma^1)\,,
		&&
		\hat\gamma^3
		=\mqty( 0&\sigma^1 \\ -\sigma^1&0)\,,
	\end{align}
	and the resulting $(2+1)$-dimensional equations coincide with Eqs.~\eqref{eq:Gammaup}--\eqref{eq:Gammadown}.
	
	The algebraic background of this discussion  as well as its generalization to arbitrary dimensions will be the topic of future reserach.

	\subsubsection{Conserved \texorpdfstring{$\kappa_3$}{k3}-charge.}
	
	An interesting consequence of the descent is that the transformation 
	\begin{equation}\label{eq:sym}
		\Psi(x)\to \mathrm{e}^{\iu\theta \kappa^3}\Psi(x)
	\end{equation}
	becomes a symmetry of the reduced Lagrangian~\eqref{eq:lagrDesc}, with Noether current 
	\begin{equation}
		j_{\kappa^3}^\mu=\overline\Psi\gamma^\mu \kappa^3\Psi\,.
	\end{equation}
	We stress that the transformation~\eqref{eq:sym} is not a symmetry of the Dirac Lagrangian~\eqref{eq:lagrDirac}, as $\kappa^3$ does not commute (in fact, it anticommutes) with $\gamma^3$:
	\begin{equation}\label{eq:Pgamma3}
		\comm{\gamma^3}{\kappa^3}=-2\gamma^5\,, \qquad \acomm{\gamma^3}{\kappa^3}=0\,.
	\end{equation}
	To better understand the physical meaning of the symmetry~\eqref{eq:sym}, it is useful to evaluate the Noether charge 
	\begin{equation}
		Q_{\kappa^3}=\int j^0_{\kappa^3}\, \dd^2{\vb*x}=\int\left(\abs{\Psi_+}^2-\abs{\Psi_-}^2\right) \dd^2{\vb*x},
	\end{equation}
	whose conservation implies that, after the descent, the current $I_+=\partial_0 \int \abs{\Psi_+}^2 \dd^2{\vb*x}$ is always balanced by $I_-=\partial_0 \int \abs{\Psi_-}^2 \dd^2{\vb*x}$. The theory is split in two sectors with  superselection charge $\kappa_3=P_+-P_-$.

	\subsubsection{Analogy with the massless theory.} 
	
	We can draw a somewhat broad parallel between the reduced Dirac theory and the massless case. When $m=0$, the Dirac equation splits into two uncoupled Weyl equations, a left- and a right-handed one, in terms of the left- and the right-handed part of the wavefunction, which are eigenvectors of $\gamma^5$ corresponding to the eigenvalues $1$ and $-1$ respectively. This decomposition has a restricted covariance, and is manifest in the Weyl representation. Moreover the chiral tran\-sfor\-ma\-tion $\Psi\to \mathrm{e}^{\iu\theta \gamma^5}\Psi$ is a symmetry of the massless Dirac Lagrangian~\eqref{eq:lagrDirac}, associated to the axial current $\overline{\Psi}\gamma^\mu\gamma^5\Psi$. The roles of mass, chirality, and the Weyl representation for the massless case are played by the $z$-component of the momentum, the superselection charge $\kappa^3$, and any block-diagonal representation in the reduced theory, respectively.
	
	\section{Dimensional reduction of the minimally coupled Dirac theory}\label{sec:diracem}
	
	\subsection{Descent of the Maxwell equations}\label{sec:freeem} 
	Let us recall the basics of electromagnetic (EM) theory in presence of non-dynamical sources. In the field formulation, the Maxwell equations,
	\begin{numcases}{}\label{eq:MaxwellHom}
		\epsilon^{\mu\nu\rho\sigma}	\partial_{\nu}F_{\rho\sigma}=0\,,\\
		\partial_{\nu}F^{\nu\mu}=j^{\mu}\,,            \label{eq:MaxwellInhom}
	\end{numcases}
	are a system of eight first-order differential equations for the components of the EM field tensor $(F^{\mu\nu})$, with the source terms corresponding to the components of the four-current $(j^{\mu})$:
	\begin{equation}
		\label{eq:Fj}
		(F^{\mu\nu})=
		\left(\begin{array}{@{}cccc@{}}
			0 & -E_x & -E_y & -E_z\\
			E_x & 0 & -B_z & B_y\\
			E_y & B_z & 0 &-B_x\\
			E_z & -B_y&B_x  & 0
		\end{array}\right),
		\qquad
		(j^\mu)=	\left(\begin{array}{@{}c@{}}
			\rho\\
			J_x\\
			J_y\\
			J_z
		\end{array}\right).
	\end{equation}
	In the potential formulation, the homogeneous equations~\eqref{eq:MaxwellHom} are equivalent to the existence of a four-potential $(A^\mu)=(\varPhi, \vb*A )$, determining the field components via 
	\begin{equation}\label{eq:MaxIden}
		F^{\mu\nu}=\partial^\mu A^\nu-\partial^\nu A^\mu\,,
	\end{equation}
	and the inhomogeneous equations~\eqref{eq:MaxwellInhom} are rewritten as a system of four second-order differential equations in the $4$-potential components, 
	\begin{equation}\label{eq:poten}
		\partial_\rho(\partial^\rho A^\mu-\partial^\mu A^\rho)=j^\mu\,,	
	\end{equation}
	that can be obtained as the Euler-Lagrange equations of the EM Lagrangian
	\begin{align}\label{eq:lagrEM}
		\mathcal{L}_{\mathrm{em}}
		&=-\tfrac{1}{4}F_{\mu\nu}F^{\mu\nu}-J_\mu A^\mu
		=\tfrac{1}{2}\qty(\vb*{E}^2-\vb*{B}^2)-\rho\,\varPhi+\vb*J\cdot\vb*A\,,
	\end{align}
	which is invariant with respect to gauge transformations
	\begin{align}
		A^\mu(x)\to A^\mu(x) -\partial^\mu\chi(x)\,.
	\end{align}	
	
	The dimensional reduction of EM has been extensively discussed in~\cite{descent}. The descent from $3$ to $2$ spatial dimensions of the Maxwell equations~\eqref{eq:MaxwellHom}--\eqref{eq:MaxwellInhom} is quite similar to the one of the free Dirac equation of Section~\ref{sec:freedirac}: chosen $z$ as the descent direction, Maxwell's differential problem is specialized into a $z$-in\-de\-pen\-dent one by imposing the conditions
	\begin{equation}\label{eq:de3EM}
		\partial_3F^{\mu\nu}=0\,, \qquad \partial_3 j^{\mu}=0\,, 
	\end{equation}
	which are covariant with respect to the $\mathrm{O}(1,2)$ group~\eqref{eq:lambda}.
	As a consequence, the original system splits up into two decoupled subsystems of four equations:
	\begin{align}
		&	\begin{cases}
			\epsilon^{b c d} \partial_{b} F_{c d}=0\,, 			
			\\		
			\partial_{b}F^{ba}=j^{a},								\label{eq:Fab}
		\end{cases}\\
		&\begin{cases}
			\epsilon^{a b c} \partial_{b} {F_{c}}^3=0, 			
			\\	
			\partial_{b}F^{b 3}=j^3\,,							\label{eq:Fa3}
		\end{cases}
	\end{align}
	and both the EM field tensor and the $4$-current are partitioned into two mutually independent sectors, according to the $\mathrm{O}(1,2)$ decomposition:
	\begin{align}
		\label{eq: 2+1decomposition}
		(F^{\mu\nu})=
		\left(\begin{array}{@{}ccc|c@{}}
			0 & -E_x & -E_y & -E_z\\
			E_x & 0 & -B_z & B_y \\
			E_y & B_z & 0 & -B_x\\
			\hline
			E_z & -B_y& B_x  & 0		\end{array}\right),
		&&
		(j^\mu)=	\left(\begin{array}{@{}c@{}}
			\rho\\ J_x\\ J_y\\ \hline J_z
		\end{array}\right).
	\end{align}	
	As the two Dirac equations, these two subsystems describe independent EM worlds, each evolving on its own. We label them according to the number of electric and magnetic components involved:
	\begin{itemize}
		\item the ``EEB'' equations~\eqref{eq:Fab} are written in terms of the field components $(E_x,E_y,B_z)$, determining the tensor block $(F^{ab})$, and of the sources $(\rho,J_x,J_y)$, determining the vector block $(j^a)$;
		\item the ``BBE'' equations~\eqref{eq:Fa3} involve the components $(B_x,B_y,E_z)$, determining the vector block $(F^{a3})$, and the source $j^3=J_z$.
	\end{itemize}
	This splitting is analogue to the one occurring in the description of the stationary EM phenomena, obeying the two independent theories of electrostatics and magnetostatics: while the descent~\eqref{eq:de3EM} consists in assuming all the EM quantities to be uniform along a spatial direction, the stationary equations are obtained by requiring uniformity along the time direction.
	
	Since the descent is performed on the equations of motion of the fields, the decoupling at the level of the potentials is not as manifest as in the field formulation. The descent conditions~\eqref{eq:de3EM}
	actually do \textit{not} imply $\partial_3A^{\mu}=0$, and the relations in Eq.~\eqref{eq:poten} specialize into
	\begin{numcases}{}
		\partial_{a}\partial^{a}A^{b}-\partial^{b}\partial_{a}A^{a}= j^{b}\,,   \label{eq:pota}\\
		\partial_{a}\partial^{a}A^3-\partial_3\partial_{a}A^{a}=j^3\,.   \label{eq:pot3}
	\end{numcases}		
	The EEB subsystem translates into Eq.~\eqref{eq:pota}, which involve no $z$ derivative term, and determine a $(2+1)$-dimensional problem, in terms of the $(2+1)$-dimensional vectors $(A^a)$ and $(j^a)$ only, that is, of the components of the four-potential and of the four-current ``transversal to the descent direction''. The BBE subsystem translates into Eq.~\eqref{eq:pot3}, involving not only the components $A^3$ and $j^3$ ``along the descent direction'', but also $(A^a)$, via the term $\partial_3 \partial_b  A^b$, which is also the only surviving $z$-derivative.	
	Therefore, while the components $(A^a)$  evolve on their own, the evolution of $A^3$ is affected by the value of $(A^a)$, as opposed to what happens at the level of the fields, where $(F^{ab})$ and $(F^{a3})$ evolve irrespectively of each other.

	In order to decouple the potential components into two independent theories, part of the gauge freedom must be used to \emph{impose} $\partial_3 A^{\mu}=0$. Then, $(F^{a3})$ is determined only by $A^3$, and Eq.~\eqref{eq:pot3} reduces to $\partial_a\partial^a A^3=j^3$, namely a $(2+1)$-dimensional wave equation for $A_3$. This choice is analogous, in the context of stationary EM, to describing fields in terms of a time-independent four-potential, decomposing into a scalar and purely electrostatic component, and a vector and purely magnetostatic one.
	
	\subsection{Descent of the coupled Dirac-Maxwell equations}
	The Lagrangian associated with a Dirac fermion, of mass $m$ and electric charge $q$, interacting with an EM field, is the sum of the free EM Lagrangian~\eqref{eq:lagrEM}, and a modified Dirac Lagrangian~\eqref{eq:lagrDirac}, where, according to the minimal coupling prescription, $p^\mu\to p^\mu-q A^\mu$, the ordinary derivatives are replaced by the covariant ones: 
	\begin{equation}
		\partial_\mu\to D_\mu=\partial_\mu+\iu qA_\mu .
	\end{equation}
	This requirement promotes the $\mathrm{U}(1)$ invariance of the free Dirac theory to a gauge symmetry:
	\begin{align}
		\Psi(x)\to\mathrm{e}^{\iu q\chi(x)}\Psi(x) , \qquad A_\mu(x)\to A_\mu(x) -\partial_\mu\chi(x) . 
	\end{align}
	Therefore, the resulting Lagrangian is the sum of the free Lagrangians~\eqref{eq:lagrDirac} and~\eqref{eq:lagrEM}, with an additional interaction term coupling the EM four-potential and the fermionic current~\eqref{eq:jmuvett}:
	\begin{align}
		\mathcal L_{\mathrm{mc}}
		&=-\tfrac{1}{4}F_{\mu\nu}F^{\mu\nu}+
		\overline \Psi\qty(\iu\gamma^\mu \partial_\mu-m)\Psi
		-qA_\mu\overline \Psi \gamma^\mu \Psi\,.
	\end{align}
	The corresponding Euler-Lagrange equations, which will be referred to as the Dirac-Maxwell equations, read
\begin{eqnarray}
		(\iu\gamma^\mu \partial_\mu-m)\Psi
		&=&qA_\mu\gamma^\mu \Psi\,,				
		\label{eq:dirMax} \\
		\partial_\mu\partial^\mu A^\nu-\partial^\nu \partial_\mu A^\mu
		&=& q\overline \Psi \gamma^\nu \Psi\,. 								
		\label{eq:maxDir}
\end{eqnarray}
	This is a system of four first-order and four second-order partial differential equations in the four complex components of $\Psi$ and the four real components of $A^\mu$, coupled to each other. Each equation has a nonlinear term of order zero, which has been written on the right-hand side, to suggest its interpretation as a dynamical source. Equation~\eqref{eq:dirMax} is a Dirac equation minimally coupled to a dynamical EM field, whereas Eq.~\eqref{eq:maxDir} represents the dynamics of an EM field with a dynamical spinor current.

	We are now going to discuss the dimensional reduction of the Dirac-Maxwell equations. We require that the reduced theory retains gauge-covariance (at the level of the equations of motion). In this regard, while the descent condition for the free EM theory, $\partial_3 F^{\mu\nu}=0$, is covariant, the one for the free Dirac theory, $\partial_3\Psi=0$, is not. We overcome this problem by generalizing the latter to the gauge-covariant condition $D_3\Psi=0$. As a consequence, the spinor is constrained to the form
	\begin{equation}
		\Psi(t,x,y,z)=\exp\left(\iu q\int^z \!A^3(t,x,y,\zeta)\,\dd{\zeta} \right) \Phi(t,x,y)\,,
	\end{equation}
	with $\Phi$ any four-component $z$-independent function. Notice that a dependence of $\Psi$ on $z$ is still allowed, as long as it is limited to a phase factor. This implies that any bilinear quantity $\Psi^\dagger M\Psi$, with $M$ a $4\times 4$ matrix, is $z$-independent:
	\begin{equation}
		\partial_3\left(\Psi^\dag M\Psi\right)=0 .
		\label{eq:physconstr}
	\end{equation}
	In particular, the spinor current is $z$-in\-de\-pen\-dent, consistently with the descent condition~\eqref{eq:de3EM} for an external four-current. In fact, the converse holds too:	it is easy to prove that the constraint~\eqref{eq:physconstr} on physical bilinear observables implies the gauge-covariant constraint on the spinor $D_3\Psi=0$.
	
	Summing up, we specialize the Dirac-Maxwell differential problem to a $z$-in\-de\-pen\-dent one by imposing the re\-pre\-sen\-ta\-tion-inde\-pen\-dent, Lo\-rentz-co\-va\-riant, and gauge-co\-va\-riant descent conditions:
	\begin{equation}\label{eq:D3Psi}
		D_3\Psi=0 , \qquad
		\partial_3\qty(\partial^\mu A^\nu-\partial^\nu A^\mu)=0 .
	\end{equation}
	Observe that the first equation represents a generalization of Hadamard's method. Each condition pertains to just one of the Dirac-Maxwell equations~\eqref{eq:dirMax}--\eqref{eq:maxDir}, hence their respective descents will be discussed one at a time.
	
	The Dirac equation~\eqref{eq:dirMax} is reduced to the equation
	\begin{equation}\label{eq:DirRedCova}
		\qty(\iu\gamma^a D_a-m\id_4)\Psi=0\,,
	\end{equation}
	which differs from the free Eq.~\eqref{eq:descent} only by the replacement of the partial derivatives with the covariant ones: the fermion is now minimally coupled to the dynamical vector potential $(A^a)$, whose components appear in the covariant derivatives $D_a$, namely to the $(2+1)$-dimensional EEB model. Using the resolution of identity provided by the projections~\eqref{eq:Ppm}, Eq.~\eqref{eq:DirRedCova} can be rewritten as
	\begin{align}
		\qty(\iu\gamma_{+}^{a}D_a-m\id_4)\Psi_{+}&=0\,,			\label{eq:DirPcoup}\\
		\qty(\iu\gamma_{-}^{a}D_a-m\id_4)\Psi_{-}&=0\,,			\label{eq:DirMcoup}
	\end{align}
	which generalizes the pair~\eqref{eq:gammap}--\eqref{eq:gammam} of $(2+1)$-dimensional free Dirac equations: the two fermionic sectors in $2+1$ dimensions are now minimally coupled to the same low-dimensional EM field.
	
	The Maxwell equations~\eqref{eq:maxDir} are reduced to
	\begin{align}
		\partial_a F^{ab}
		&= q \overline \Psi\gamma^{b} \Psi\,,										\label{eq:DescMaxDir012}\\
		\partial_a F^{a3}
		&=q \overline \Psi\gamma^3\Psi\,,														\label{eq:DescMaxDir3}
	\end{align}
	which are respectively equal to Eq.~\eqref{eq:pota}, for the EEB model, and Eq.~\eqref{eq:pot3}, for the BBE model, except that the external $(2+1)$-dimensional vector and scalar currents $(j^{ a})$ and $j^3$ have been replaced respectively by the $(2+1)$-dimensional blocks $(q\overline \Psi\gamma^{ a} \Psi)$ and $q\overline \Psi\gamma^3\Psi$ of the dynamical spinor current.
	
	Summarizing, the reduced Dirac-Maxwell equations can be finally written as
	\begin{equation}
		\begin{cases}
			(\iu\gamma^a \partial_a-m\id_4)\Psi=qA_a\gamma^a \Psi\,,				
			\\
			\partial_{a}(\partial^{a}A^{b}-\partial^{b}A^{a})
			= q\overline \Psi \gamma^b \Psi\,, 										
			\\
			\partial_{a}(\partial^{a}A^{3}-\partial_{3}A^{a})
			= q\overline \Psi \gamma^3 \Psi\,.
			\label{eq:maxDir3}
		\end{cases}
	\end{equation}
	in terms of $\Psi$, $(A^a)$ and $A^3$, or, equivalently, by using the decompositions~\eqref{eq:Decomp},
	\begin{equation}
		\begin{cases}
			\qty(\iu\gamma_{+}^{a}\partial_a-m\id_4)\Psi_{+}=qA_a\gamma_+^a \Psi_+\,,			
			\\
			\qty(\iu\gamma_{-}^{a}\partial_a-m\id_4)\Psi_{-}=qA_a\gamma_-^a \Psi_-\,,			
			\\
			\partial_{a}(\partial^{a}A^{b}-\partial^{b}A^{a})
			=j^{b}_++j^{b}_-\,,							
			\\
			\partial_{a}(\partial^{a}A^{3}-\partial_{3}A^{a})
			=j^3_{+,-}+j^3_{-,+}\,.	
		\end{cases}
	\end{equation}
	Furthermore, by a block-diagonal choice~\eqref{eq:diagonal_decomp} for the gamma matrices, the Dirac-Maxwell equations can be recast in the form
	\begin{numcases}{}
		\bigl(\iu\hat\Gamma_{+}^{a}\partial_a-m\id_2\bigr)\hat\psi_{+}=qA_a\hat\Gamma_{+}^{a} \hat\psi_+\,,			\label{eq:dirMaxP}\\
		\bigl(\iu\hat\Gamma_{-}^{a}\partial_a-m\id_2\bigr)\hat\psi_{-}=qA_a\hat\Gamma_{-}^{a} \hat\psi_-\,,			\label{eq:dirMaxM}\\
		\partial_{a}\partial^{a}A^{b}-\partial^{b}\partial_{a}A^{a}
		=j^{b}_++j^{b}_-\,,							\label{eq:EMa} \\
		\partial_{a}\partial^{a}A^{3}+\partial_{3}\partial_{a}A^{a}
		=j^3_{+,-}+j^3_{-,+}\,.	\label{eq:EM3}
	\end{numcases}
	in which the number of independent components is explicit.
	
	\begin{figure}[tp]
		\centering
		\includegraphics[width=0.8\textwidth]{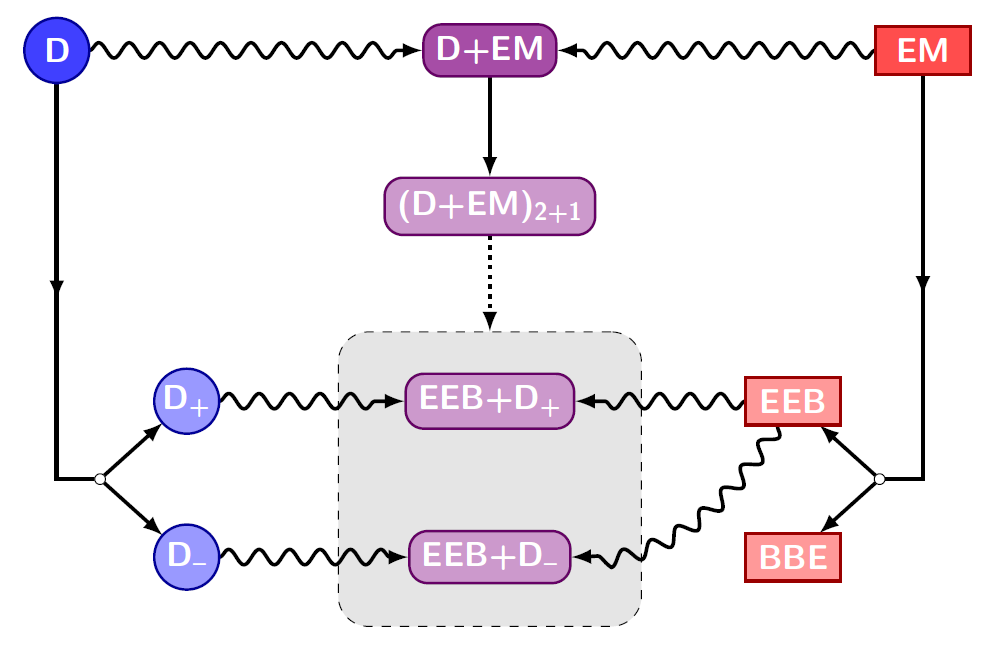}
		\caption{Relationship between descent and minimal coupling. The solid straight lines represent the descent from a $3+1$ theory to a $2+1$ one, the solid wavy lines represent the minimal coupling, whereas the dotted line represents the restriction of a theory to a particular class of solutions.}
		\label{fig-schema}
	\end{figure}
	
	\subsection{Discussion of the results}
	
	As opposed to what happens in the free theories, the Dirac-Maxwell equations do \textit{not} split by descent into two separate, non-interacting sectors, as variables are directly or indirectly coupled to each other. In particular, $\Psi_{+}$ and $\Psi_{-}$ are coupled between them through the EM field. Decoupling is only partial, with Eqs.~\eqref{eq:dirMaxP}--\eqref{eq:EMa} forming a subsystem in which only $\Psi$ and $(A^a)$ explicitly appear, although $A^3$ is still coupled to $(A^a)$ via the term $\partial_{3}\partial_{a}A^{a}$ (which retains a residual $z$-dependence allowed by gauge invariance), and to $\Psi$ via $j^3$.
	
	However, one can identify two classes of solutions in which decoupling effectively occurs throughout the evolution:
	\begin{enumerate}
		\item The solutions in which $\Psi_{-}$ identically vanishes, and the remaining fields consistently evolve according to
		\begin{equation}
			\begin{cases}
				\qty(\iu\gamma_{+}^{a}D_a-m \id_4)\Psi_{+}=0\,,
				\\
				\partial_{a}F^{ab}=j^{b}_+
				\\
				\partial_{a}F^{a3}=0\,.	
			\end{cases}
		\end{equation}
		\item The solutions in which $\Psi_{+}$ identically vanishes, and the remaining fields consistently evolve according to
		\begin{equation}
			\begin{cases}
				\qty(\iu\gamma_{-}^{a}D_a-m \id_4)\Psi_{-}=0\,,
				\\
				\partial_{a}F^{ab}=j^{b}_-
				\\
				\partial_{a}F^{a3}=0\,.	
			\end{cases}
		\end{equation}
	\end{enumerate}
	Notice that, in order to obtain the above classes of solutions, it is sufficient to impose the initial conditions $\Psi_-(t=0)=0$ and $\Psi_+(t=0)=0$, respectively. We remind that the above set of equations can be used to describe the physics of graphene interacting with an electromagnetic field \cite{graphene} [see comments on Eqs.~\eqref{eq:Gammaup}--\eqref{eq:Gammadown} in Section~\ref{sec:sr}].
	
	Formally, both these theories describe a two-com\-po\-nent spinor in $2+1$ dimensions interacting with an EEB theory, plus a non-interacting BBE sector which evolves freely. We remark that this result provides a justification \textit{a posteriori} of the traditional formulation of QED in $2+1$ dimensions, in term of two-component spinors.
	
	On the other hand, a solution with $F^{ab}$ identically vanishing would also entail, for $q\neq 0$, an identically vanishing spinor, since 
	\begin{equation}
		\partial_a F^{a3} = 0 \quad \Rightarrow \quad \Psi^{\dagger}\Psi = 0 ,
	\end{equation}
	leaving the free BBE sector as the only nontrivial theory.

	All the results discussed in this work are schematically summarized in  Fig.~\ref{fig-schema}, where the sectors D\textsubscript{+} and D\textsubscript{--} respectively refer to the free $(2+1)$-dimensional Dirac equations~\eqref{eq:Gammap} and~\eqref{eq:Gammam}, whereas the  EEB+D\textsubscript{+} and EEB+D\textsubscript{--} denote the above decoupled theories obtained when $\Psi_{-}=0$ and $\Psi_{+}=0$, respectively.

	\section{Conclusions and outlook}\label{sec:conclusions}
	
	We have shown that dimensional reduction through the descent method entails a splitting of the Dirac theory for a free spin-\textonehalf{} particle in a $(3+1)$-dimensional flat spacetime into two noninteracting sectors in a $(2+1)$-dimensional spacetime. Each sector obeys an equation that could also be obtained \textit{ab initio} as a Dirac equation in a $(2+1)$-dimensional spacetime, when defined as a formal ``square root'' of the Klein-Gordon equation. In this sense, Dirac's approach and dimensional reduction prove to be mutually compatible.
	
	We have subsequently applied the reduction procedure to a spin-\textonehalf{} particle minimally coupled to an electromagnetic field: while a complete splitting of the theory does not occur in such a case at the level of equations, two particular classes of solutions exist in which some sectors evolve trivially. In other words, the splitting occurs at the level of solutions. We emphasize that Hadamard's method has been extended by using the covariant derivative.
	
	Interestingly, the existence of \emph{two} $(2+1)$-dimensional Dirac equations after dimensional reduction is not a mere academic curiosity, since an analogous result is obtained, although under different physical conditions, in the low-energy description of the electronic properties of graphene close to the Dirac points~\cite{graphene}.
	
	We have shown that the dimensional reduction implies the emergence of a superselection charge, which is the analogous of chirality in the massless case. As it is well-known, the chiral symmetry of the $(3+1)$-dimensional massless Dirac Lagrangian is actually broken at the level of quantum field theory, a phenomenon which is usually referred to as chiral anomaly~\cite{Adler,RPBG21}. Therefore, an interesting open question is whether our newly found symmetry~\eqref{eq:sym} is still conserved in a corresponding $(2+1)$-dimensional quantum field theory.

	\section*{Acknowledgments}
	
	This research was funded by MIUR via PRIN 2017 (Progetto di Ricerca di Interesse Nazionale), project QUSHIP (2017SRNBRK), by the Italian National Group of Mathematical Physics (GNFM-INdAM), by Istituto Nazionale di Fisica Nucleare (INFN) through the project ``QUANTUM'', and by Regione Puglia and QuantERA ERA-NET Cofund in Quantum Technologies (Grant No. 731473), project QuantHEP.


\section*{References}
	\begin{thebibliography}{99}
		
			
		\bibitem{ehrenfest} P. Ehrenfest, ``In what way does it become manifest in the fundamental laws of physics that space has three dimensions?''. Proceedings of the Royal Academy of Sciences at Amsterdam \textbf{20}, 200 (1918).
		
		\bibitem{hardy} G. H. Hardy, \textit{A Mathematician's Apology}. Cambridge University Press, 2012; 1st pub.\ 1940, with foreword 1967.
				
		\bibitem{hadamard} J. Hadamard, \textit{Lectures on Cauchy's problem in linear partial differential equations}. Yale University Press, New Haven, 1923.
		
		\bibitem{evans} L. Evans, \textit{Partial Differential Equations}. American Mathematical Society Providence, 1998.
		
		\bibitem{courant} R. Courant, and D. Hilbert, \textit{Methods of Mathematical Physics, vol II}. Interscience (Wiley), New York, 1962.	
		
		\bibitem{ehrenulen} P. Ehrenfest, and G. E. Uhlenbeck, ``On the Connection of Different Methods of Solution of the Wave Equation in Multi-dimensional Spaces''. Proceedings of the Royal Academy of Sciences at Amsterdam \textbf{29}, 1280 (1926).
		
		\bibitem{balasz} N. L. Balazs, ``Wave Propagation in Even and Odd Dimensional Spaces''. Proceedings of the Physical Society Section A \textbf{68}, 521 (1955).
		
		\bibitem{qed1}
		J. S. Schwinger. ``Gauge invariance and mass. II'' Phys. Rev. \textbf{128}, 2425 (1962).
		
		\bibitem{qed2}
		J. H. Lowenstein, and J. A. Swieca, ``Quantum electrodynamics in two-dimensions'', Ann. Phys. (N.Y.) \textbf{68}, 172 (1971).
		
		\bibitem{qed3}
		S. R. Coleman, R. Jackiw, and L. Susskind, ``Charge shielding and quark confinement in the massive Schwinger model'', Ann. Phys. (N.Y.) \textbf{93}, 267 (1975).
		
		\bibitem{qed4}
		S. R. Coleman, ``More about the massive Schwinger model'' Ann. Phys. (N.Y.) \textbf{101}, 239 (1976).
		
		\bibitem{qed5}
		T. Appelquist, M. Bowick, E. Cohler, and L. C. R. Wijewardhana, ``Chiral-symmetry breaking in 2+1 dimensions'', Phys. Rev. Lett. \textbf{55}, 1715 (1985).
		
		\bibitem{qed6}
		P. Cea, ``Variational approach to (2+1)-dimensional QED'', Phys. Rev. D \textbf{32}, 2785 (1985).		
		
		\bibitem{qsim1}
		E. Zohar, J. I. Cirac, and B. Reznik, ``Simulating Compact Quantum Electrodynamics with Ultracold Atoms: Probing Confinement and Nonperturbative Effects''. Phys. Rev. Lett. \textbf{109}, 125302 (2012).
		
		\bibitem{qsim2}
		D. Banerjee, M. Dalmonte, M. M\"uller, E. Rico, P. Stebler, U. J. Wiese, and P. Zoller, ``Atomic Quantum Simulation of Dynamical Gauge Fields Coupled to Fermionic Matter: From String Breaking to Evolution after a Quench''. Phys Rev. Lett. \textbf{109}, 175302 (2012).
		
		\bibitem{qsim3}
		S. Notarnicola, E. Ercolessi, P. Facchi, G. Marmo, S. Pascazio, and F. V. Pepe,
		``Discrete Abelian Gauge Theories for Quantum Simulations of QED'',
		J. Phys. A: Math. Theor. \textbf{48}, 30FT01 (2015). 
		
		\bibitem{qsim4}
		T. Pichler, M. Dalmonte, E. Rico, P. Zoller, and S. Montangero, 
		``Real-Time Dynamics in U(1) Lattice Gauge Theories with Tensor Networks'',
		Phys. Rev. X \textbf{6}, 011023 (2016).
		
		\bibitem{qsim5}
		E. Ercolessi, P. Facchi, G. Magnifico, S. Pascazio, and F. V. Pepe, ``Phase transitions in $\mathbb{Z}_n$ gauge models: Towards quantum simulations of the Schwinger-Weyl QED''. Phys. Rev. D \textbf{98}, 074503 (2018).
		
		\bibitem{qsim6}
		G. Magnifico, M. Dalmonte, P. Facchi, S. Pascazio, F. V Pepe, and E. Ercolessi, ``Real Time Dynamics and Confinement in the $\mathbb{Z}_n$ Schwinger-Weyl lattice model for 1+1 QED''. Quantum \textbf{4}, 281 (2020).
		
		\bibitem{wqed1}
		P. Facchi, D. Lonigro, S. Pascazio, F. V. Pepe, and D. Pomarico, ``Bound states in the continuum for an array of quantum emitters''. Phys. Rev. A. \textbf{100}, 023834 (2019).
		
		\bibitem{wqed2}
		D. Lonigro, P. Facchi, S. Pascazio, F. V. Pepe, and D. Pomarico, ``Stationary excitation waves and multimerization in arrays of quantum emitters''. New J. Phys. \textbf{23}, 103033 (2021).
		
		\bibitem{roy} D. Roy, C. M. Wilson, and O. Firstenberg. ``Colloquium: Strongly interacting photons in one-dimensional continuum''. Rev. Mod. Phys. \textbf{89}, 021001  (2017).			
		
		\bibitem{graphene} A. H. Castro Neto, F. Guinea, N. M. R. Peres, K. S. Novoselov, and A. K. Geim ``The electronic properties of graphene''. Rev. Mod. Phys. \textbf{81}, 109 (2009)
		
		\bibitem{wheelerdirac} N. Wheeler, ``Dirac Equation in 2-dimensional Spacetime''. Reed College, 2000, \url{https://www.reed.edu/physics/faculty/wheeler/documents/Classical%20Field%20Theory/Miscellaneous%20Essays/A.%202D%20Dirac%20Equation.pdf}.
			
		\bibitem{qsim_book}
		M. Lewenstein, A. Sanpera, and V. Ahufinger, \textit{Ultracold Atoms in Optical Lattices: Simulating Quantum Many-Body Systems}, Oxford University Press, New York (2012).
				
		\bibitem{lapidus} I. R. Lapidus, ``Classical electrodynamics in a universe with two space dimensions''. Am. J. Physics \textbf{50}, 155 (1982).
		
		\bibitem{moreno} E. Moreno, and J. Rivas, ``On electromagnetism in $R^{2+1}$''. Eur. J. Physics \textbf{5}, 20 (1984).
		
		\bibitem{wheelerem} N.~Wheeler, ````Electrodynamics'' in 2-dimensional Spacetime''. Reed College, 1997, \url{https://www.reed.edu/physics/faculty/wheeler/documents/Electrodynamics/Miscellaneous%20Essays/E&M%20in%202%20Dimensions.pdf}.
		
		\bibitem{mcdonald} K. T. McDonald, \textit{Electrodynamics in 1 and 2 Spatial Dimensions}. Joseph Henry Laboratories, Princeton University, 2019, \url{http://kirkmcd.princeton.edu/examples/2dem.pdf}.
		
		\bibitem{boito} D. Boito, L. N. S. de Andrade, G. de Sousa, R. Gama, and C. Y. M. London, ``On Maxwell's electrodynamics in two spatial dimensions''. Revista Brasileira de Ensino de Fisica \textbf{42}, e20190323 (2020).
		
		\bibitem{GBB} \'E. Goulart, E. Bittencourt, E. O. S. Brand\~ao, ``Light propagation in (2+1)-dimensional electrodynamics: the case of linear constitutive laws", arXiv:2209.00770 [gr-qc].
		
	    \bibitem{descent} R. Maggi, E. Ercolessi, P. Facchi, G. Marmo, S. Pascazio, and F. V. Pepe, ``Dimensional reduction of electromagnetism'', J. Math. Phys. \textbf{63}, 022902 (2022). 	
							
		\bibitem{Go55} R. H. Good, ``Properties of the Dirac Matrices'', Rev. Mod. Phys. \textbf{27}, 187 (1955).
				
		\bibitem{dirac} P. A. M. Dirac, \textit{Principles of Quantum Mechanics}. International Series of Monographs on Physics (4th ed.), Oxford University Press, p. 255, 1958.
		
		\bibitem{thaller} B. Thaller, \textit{The Dirac Equation}. Springer-Verlag Berlin Heidelberg, 1992.
		
		\bibitem{BrauerWeyl} R.~Brauer, and H.~Weyl, ``Spinors in $n$ dimensions''. Am. J. Math. \textbf{57}, 425 (1935). 
	
		\bibitem{history} P. A. M. Dirac, ``Recollections of an exciting era'', in \emph{History of Twentieth-Century Physics}, Proceedings of the International School of Physics ``Enrico Fermi'', Course 57, edited by C.~Weiner, Academic Press, New York and London, 1977.		
				
%
%
		
		\bibitem{Adler} S. L. Adler, ``Axial-Vector Vertex in Spinor Electrodynamics'', Phys. Rev. \textbf{177}, 5 2426--2438 (1969). 
		
		\bibitem{RPBG21} C. Rylands, A. Parhizkar, A. A. Burkov, and V. Galitski, ``Chiral Anomaly in Interacting Condensed Matter Systems'',  Phys. Rev. Lett. \textbf{126}, 185303 (2021). 
		
		
	\end{thebibliography}
\end{document}